\documentclass[aps,pra,reprint,superscriptaddress,showpacs]{revtex4-2}

\usepackage{graphicx}
\usepackage{color}
\usepackage{url}

\begin{document}


\title{Atomistic simulation of the FEBID-driven growth of iron-based nanostructures}

\author{Alexey Prosvetov}
\email{prosvetov@mbnexplorer.com}
\affiliation{MBN Research Center, Altenh\"oferallee 3, 60438 Frankfurt am Main, Germany}

\author{Alexey V. Verkhovtsev}
\affiliation{MBN Research Center, Altenh\"oferallee 3, 60438 Frankfurt am Main, Germany}
\affiliation{On leave from the Ioffe Physical-Technical Institute, Polytekhnicheskaya 26, 194021 St. Petersburg, Russia}

\author{Gennady Sushko}
\affiliation{MBN Research Center, Altenh\"oferallee 3, 60438 Frankfurt am Main, Germany}

\author{Andrey V. Solov'yov}
\affiliation{MBN Research Center, Altenh\"oferallee 3, 60438 Frankfurt am Main, Germany}
\affiliation{On leave from the Ioffe Physical-Technical Institute, Polytekhnicheskaya 26, 194021 St. Petersburg, Russia}

\date{\today}

\begin{abstract}
The growth of iron-containing nanostructures in the process of focused electron beam-induced deposition (FEBID) of Fe(CO)$_5$ is studied by means of atomistic irradiation-driven molecular dynamics (IDMD) simulations.
The geometrical characteristics (lateral size, height and volume), morphology and metal content of the grown nanostructures are analyzed at different irradiation and precursor replenishment conditions corresponding to the electron-limited and precursor-limited regimes (ELR \& PLR) of FEBID. A significant variation of the deposit's morphology and elemental composition is observed with increasing the electron current from 1 to 4~nA. At low beam current (1~nA) corresponding to the ELR and a low degree of Fe(CO)$_5$ fragmentation, the nanogranular structures are formed which consist of isolated iron clusters embedded into an organic matrix.
In this regime, metal clusters do not coalesce with increasing electron fluence, resulting in relatively low metal content of the nanostructures.
A higher beam current of 4~nA corresponding to the PLR facilitates the precursor fragmentation and the coalescence of metal clusters into a dendrite-like structure with the size corresponding to the primary electron beam. The IDMD simulations enable atomistic-level predictions on the nanoscopic characterization of the initial phase of nanostructure growth in the FEBID process. These predictions can be verified in high-resolution transmission electron microscopy experiments.
\end{abstract}

\maketitle

\section{Introduction}
\label{sec:Intro}

Focused electron beam-induced deposition (FEBID) is a rapidly emerging technique for the direct-write fabrication of 2D and 3D nanostructures with complex geometries \cite{Utke_book_2012, DeTeresa-book2020, Winkler2018}. FEBID is based on the electron irradiation of precursor molecules (mainly organometallic) adsorbed on a substrate. Electron-induced decomposition of precursors releases the volatile, metal-free fragments pumped out of the working chamber. As a result, the non-volatile, metal-containing fragments form a deposit with a size similar to that of the incident electron beam (down to a few nanometers) \cite{Utke2008}.
In FEBID, the adsorbed precursors are irradiated by a high-energy primary electron (PE) beam with typical energy ranging from 1 to 30~keV. Decomposition of precursors is primarily induced by secondary low-energy electrons (with the energy up to several tens of eV) produced as the primary beam impinges on the substrate's surface \cite{Thorman2015}.

One of the not entirely resolved technological challenges for FEBID is the controllable nanoscale fabrication of high-purity metal nanostructures of the desired geometry, size and composition. Contamination of the metal deposit (typically with carbon and oxygen) is inherent in the FEBID process, as some of the precursor residues are easily integrated into the deposit together with the metal material \cite{Botman2009a}. Irradiation-driven chemistry plays a crucial role in determining the elemental composition and spatial resolution of the grown nanostructures. Low-energy secondary electrons (SEs) emitted outside the focal point of the PE beam initiate electron-driven reactions and induce the formation of a halo, thus producing an undesired edge broadening of the structure \cite{Arnold2014}. These limitations have motivated a dedicated experimental, theoretical and computational effort for elucidating the underlying mechanisms of FEBID, which determine the elemental composition and morphology of the grown deposits.

FEBID operates through successive cycles of precursor molecules replenishment on a substrate and irradiation by a focused electron beam \cite{Huth2012}. It involves a complex interplay of phenomena taking place on different temporal and spatial scales: (i) adsorption, diffusion and desorption of precursors on the substrate; (ii) transport of PEs, SEs and backscattered electrons (BSEs); (iii) electron-induced dissociation of the adsorbed molecules; and (iv) the follow-up chemistry. All these phenomena can be explored in detail using an advanced multiscale modeling approach. Successful examples of the multiscale computational modeling of FEBID have been reviewed in Refs.~\cite{MBNbook_Springer_2017, Verkhovtsev_EPJD2021_MM}.

The atomistic modeling of FEBID has become possible recently through Irradiation-Driven Molecular Dynamics (IDMD) \cite{Sushko2016}, a novel and general methodology for computer simulations of irradiation-driven transformations of complex molecular systems. In contrast to other commonly used computational methods for studying FEBID, based on the Monte Carlo approach and the diffusion-reaction theory \cite{Utke_book_2012, Fowlkes2010, Sanz-Hernandez2017, Toth2015}, IDMD enables the atomistic-level description of the FEBID nanostructures growth by accounting for irradiation-induced chemical transformation of surface-adsorbed molecular systems under focused electron beam irradiation \cite{Sushko2016, MBNbook_Springer_2017, DeVera2020}.

Within the IDMD framework \cite{Sushko2016}, various quantum processes occurring in an irradiated system (e.g. ionization, bond dissociation via electron attachment, or charge transfer) are treated as random, fast and local transformations incorporated into the classical MD framework in a stochastic manner. The probabilities of these quantum processes are elaborated through the corresponding cross sections.
Major transformations of irradiated molecular systems (such as change in molecular topology, change of interatomic interactions, or redistribution of atomic partial charges) are simulated by means of MD with the reactive rCHARMM force fields \cite{Sushko2016a} using MBN Explorer \cite{Solovyov2012} -- a multi-purpose software package for multiscale simulations of the structure and dynamics of complex Meso-Bio-Nano (MBN) systems \cite{MBNbook_Springer_2017}. The IDMD simulation outcomes are analyzed by means of MBN Studio \cite{Sushko2019} -- a powerful multitask toolkit for MBN Explorer, enabling the fast and efficient computational design and characterization of various MBN systems.

In Ref.~\cite{Sushko2016} IDMD was applied for the first time to simulate FEBID of W(CO)$_6$ precursor molecules on a SiO$_2$ surface and enabled to predict the molecular composition and growth rate of tungsten-based nanostructures emerging on the surface during the FEBID process.
The follow-up study \cite{DeVera2020} introduced a novel multiscale computational methodology that couples track-structure Monte Carlo simulations for electron transport with IDMD for atomistic simulations of the irradiation-driven chemistry processes during FEBID. The spatial and energy distributions of SEs and BSEs emitted from a SiO$_2$ substrate were used to simulate the electron-induced nanostructure formation and growth considering W(CO)$_6$ precursors as a case study. The cited studies demonstrated that IDMD simulations provide insights into the deposits' internal structure and its evolution depending on the irradiation and replenishment regimes of the FEBID process. The IDMD approach also enables monitoring and predicting the morphology of the FEBID nanostructures on the atomistic level \cite{Sushko2016, DeVera2020}.

Our recent study~\cite{Prosvetov2021} provided a detailed description of the IDMD-based computational methodology for modeling the formation and growth of FEBID nanostructures. Different computational aspects of the methodology and the key input parameters describing the precursor molecules, the substrate, and the irradiation and replenishment conditions were systematically described. The formulated computational protocol was applied to simulate the FEBID of Pt(PF$_3$)$_4$ precursor molecules on a fully hydroxylated SiO$_2$ (SiO$_2$-H) surface. Particular focus was made on the atomistic characterization of the initial stage of the FEBID process, including nucleation of platinum atoms, formation of small metal clusters on the surface followed by their aggregation, and, eventually, the formation of dendritic platinum nanostructures.
The IDMD simulations carried out in Ref.~\cite{Prosvetov2021} revealed a morphological transition from isolated metal-enriched islands formed on the substrate into a single metal nanostructure.
The morphology of deposits governs many physical properties such as electrical and thermal conductivity and magnetic properties \cite{Huth_2009, Porrati2011, Huth2020}.

The characteristics of the FEBID nanostructures (such as lateral size, height, metal concentration, growth rate) are determined by the precursor fragmentation rate (which depends on the energy and flux of PEs, BSEs and SEs) and the surface density of adsorbed precursors. Two different FEBID regimes can be distinguished -- the electron-limited regime (ELR) and the precursor-limited regime (PLR) \cite{DeTeresa-book2020, Barth2020_JMaterChemC}. In the ELR, only a fraction of all the available precursor molecules on the surface exposed to irradiation by the PEs, BSEs and SEs become fragmented during the irradiation phase. In the PLR, all the precursor molecules available on the surface become fragmented during the specific irradiation period. In this case, a high degree of dissociation of the molecules and follow-up irradiation-driven chemical reactions determine the properties of the growing nanostructure.

The electron-limited and precursor-limited FEBID regimes have been previously studied theoretically by means of the continuum (diffusion-reaction) models \cite{Szkudlarek_ApplPhysA_2014, Toth2015, Cullen_JPCC_2015}. Experimentally, these regimes have been identified by measuring the variation of the deposits' volume as a function of beam current \cite{Wachter2014, Lavrijsen2011}. According to the experiments \cite{Wachter2014} carried out with Co$_2$(CO)$_8$ precursors, the deposit's volume increases linearly at low electron current (below approx. 0.5~nA), and the deposit's growth slows down at higher currents, indicating a transition from the ELR to the PLR. In Ref.~\cite{Lavrijsen2011} devoted to the FEBID of Fe$_2$(CO)$_9$ precursors, a linear increase in the volume growth rate (indicative of the ELR) has been observed for much higher electron beam currents up to 22~nA. The results of these studies indicate that the regime in which the FEBID process operates depends on the choice of precursor molecule as well as on irradiation and replenishment conditions of each particular experiment.

The present study explores how the morphology and composition of metal-based nanostructures vary at different irradiation and replenishment conditions corresponding to the ELR and PLR of the FEBID process. The FEBID of Fe(CO)$_5$ precursors on a SiO$_2$-H substrate is considered as an illustrative case study. Fe(CO)$_5$ is one of the most common FEBID precursors used to fabricate magnetic nanostructures for magnetic sensing, spintronics, and magnetologic technologies \cite{Takeguchi2005, Serrano-Ramon2011, Gavagnin2014, Lukasczyk2008, Gavagnin2013, Porrati2011}.
Apart from that, a large number of gas-phase and surface science experiments have been performed over the last years to study the electron-induced dissociation mechanisms for isolated Fe(CO)$_5$ molecules \cite{Lacko_EPJD2015, Allan_PCCP2018, Ribar_EPJD2015}, those embedded in a cluster environment \cite{Lengyel2016_1, Lengyel2016_2, Lengyel2017_Beilstein, Lengyel2021}, and condensed on a surface in the form of Fe(CO)$_5$ thin films \cite{Massey_Sanche2015, Bilgilisoy_Fairbrother2021}.

In the present study, the dependence of FEBID nanostructures’ morphology and metal content on the number of adsorbed molecules and electron flux is investigated on the atomistic level by means of IDMD simulations. A significant variation of the deposit's morphology and elemental composition has been observed by increasing the electron current from 1 to 4~nA, which corresponds to the transition from ELR to PLR for a given dwell time $\tau_d = 10$~ns. Nanogranular deposits are formed at low beam current (1~nA) corresponding to a small degree of Fe(CO)$_5$ fragmentation; such deposits consist of small-size iron clusters surrounded by organic ligands. In this regime, metal clusters do not agglomerate with increasing electron fluence, limiting the nanostructures' metal content by ca. 20~at.~\%.
At higher beam currents (2 and 4~nA), a higher electron flux density increases the degree of precursor fragmentation. This enables a morphological transition corresponding to the coalescence of isolated metal clusters into a single dendrite-like structure with the size corresponding to the primary electron beam. In this regime, the nanostructure's metal content increases twofold compared to the case of low current. The results of atomistic IDMD simulations reported in the present study enable predictions concerning the nanoscopic characterization of the initial phase of nanostructure growth under focused electron beam irradiation.

\section{Computational methodology}
\label{sec:Methodology}

Computer simulations of the FEBID process of Fe(CO)$_5$ have been performed by means of the MBN Explorer software package \cite{Solovyov2012}. The MBN Studio toolkit \cite{Sushko2019} has been utilized to create the systems, prepare all necessary input files and analyze simulation outputs. The step-by-step protocol of the multiscale IDMD-based simulation of the FEBID process \cite{Sushko2016, DeVera2020} has been described in detail in the earlier study \cite{Prosvetov2021} and is therefore only briefly recapped below.

Interatomic interactions involving the precursor molecules are described using the reactive CHARMM (rCHARMM) force field \cite{Sushko2016a}. rCHARMM permits simulating various molecular systems with the dynamically changing molecular topology \cite{Verkhovtsev_2017_EPJD.71.212, DeVera2019, Friis2020}, which is essential for modeling the precursor fragmentation and the formation of metal-containing nanostructures. A detailed description of rCHARMM is given in Ref.~\cite{Sushko2016a}, see also a recent review \cite{Verkhovtsev_EPJD2021_IDMD}.

The radial part of the covalent bond interactions is described in rCHARMM by means of the Morse potential:
\begin{equation}
U^{{\rm bond}}(r_{ij}) = D_{ij} \left[ e^{-2\beta_{ij}(r_{ij} - r_0)} - 2e^{-\beta_{ij}(r_{ij} - r_0)} \right] \ .
\label{Eq. Morse}
\end{equation}
Here $D_{ij}$ is the dissociation energy of the bond between atoms $i$ and $j$, $r_0$ is the equilibrium bond length, and the parameter $\beta_{ij} = \sqrt{k_{ij}^{r} / D_{ij}}$ (where $k_{ij}^{r}$ is the bond force constant) determines the steepness of the potential. The bonded interactions are truncated at a user-defined cutoff distance beyond which the bond is considered broken and the molecular topology of the system changes. The bond energy given by Eq.~(\ref{Eq. Morse}) asymptotically approaches zero at large interatomic distances.

The rupture of covalent bonds in the course of simulation employs the following reactive potential for valence angles \cite{Sushko2016a}:
\begin{equation}
U^{{\rm angle}}(\theta_{ijk}) =
2 k^\theta_{ijk} \, \sigma(r_{ij}) \, \sigma(r_{jk}) \left[ 1 - \cos(\theta_{ijk}-\theta_0 )  \right] \ ,
\label{Eq. Angles}
\end{equation}
where $\theta_0$ is the equilibrium angle formed by atoms $i$, $j$ and $k$; $k^{\theta}$ is the angle force constant; and the function
\begin{equation}
\sigma(r_{ij}) = \frac{1}{2} \left[1-\tanh(\beta_{ij}(r_{ij}-r_{ij}^*))  \right]
\label{Eq. Rupture_param}
\end{equation}
describes the effect of bond breakage. Here $r_{ij}^*=(R^{{\rm vdW}}_{ij}+r_0)/2$ with $r_0$ being the equilibrium distance between two atoms involved in the angular interaction and $R^{{\rm vdW}}_{ij}$ being the sum of the van der Waals radii for those atoms.

Two different types of metal-ligand bonds can be distinguished in a Fe(CO)$_5$ molecule: two axial (``ax'') CO groups lie on the main symmetry axis of the molecule, while three equatorial (``eq'') CO groups lie in the plane perpendicular to the main axis.

The initial geometry of a Fe(CO)$_5$ molecule has been determined via density-functional theory (DFT) calculations using Gaussian 09 software \cite{Gaussian09} and then optimized using MBN Explorer \cite{Solovyov2012}. The rCHARMM parameters for a Fe(CO)$_5$ molecule have been evaluated from a series of DFT-based potential energy scans, following the protocol employed in the earlier studies~\cite{DeVera2019, Prosvetov2021} for W(CO)$_6$ and Pt(PF$_3$)$_4$ precursors. The parameters of the bonded and angular interactions for Fe(CO)$_5$ are listed in Table~\ref{Table:CovBonds}. In agreement with Ref.~\cite{Lacko_EPJD2015}, the data listed in Table~\ref{Table:CovBonds} indicate that the dissociation energy for Fe--C$_{\rm eq}$ bonds is lower than that for Fe--C$_{\rm ax}$ bonds despite the close values of the equilibrium bond lengths.

\begin{table*}[t!]
\centering
\caption{Covalent bonded and angular interaction parameters for a Fe(CO)$_5$ molecule employed in the simulations.}
\begin{tabular}{p{3.5cm}p{2cm}p{3.8cm}p{3.5cm}}
\hline
bond type      &   $r_0$~(\AA)   &  $D_{ij}$~(kcal/mol)  &  $k_{ij}^{r}$~(kcal/mol \AA$^{-2}$)  \\
\hline
Fe -- C$_{\rm ax}$       &    1.88     &    37.1   &   111.3  \\
Fe -- C$_{\rm eq}$       &    1.90     &    25.6   &    78.2  \\
C$_{\rm ax / eq }$ -- O  &    1.12     &   227.6   &  1564.3  \\
\hline
\hline
angle type  &  $\theta_0$~(deg.)  & $k_{ijk}^{\rm {\theta}}$~(kcal/mol rad$^{-2}$)   \\
\hline
C$_{\rm ax}$ -- Fe -- C$_{\rm ax}$   &  180   &  76.4  \\
C$_{\rm eq}$ -- Fe -- C$_{\rm eq}$   &  120   &  76.4  \\
C$_{\rm ax}$ -- Fe -- C$_{\rm eq}$   &   90   &  76.4  \\
Fe -- C$_{\rm ax / eq}$ -- O         &  180   &  28.0  \\
\hline
\end{tabular}
\label{Table:CovBonds}
\end{table*}

The interaction between iron atoms in the formed metal-containing structures has been described by means of the many-body embedded-atom-model (EAM) potential \cite{Ackland_EAM_Fe} taken from the NIST Interatomic Potentials Repository\footnote{https://www.ctcms.nist.gov/potentials/}.

Following the earlier IDMD-based studies of FEBID \cite{Sushko2016,DeVera2020, Prosvetov2021}, the SiO$_2$-H substrate has been fixed in space in the course of simulations for computational speed-up.
The adsorbed precursor molecules and fragments interact with the substrate via van der Waals forces described by means of the Lennard-Jones potential:
\begin{equation}
U_{{\rm LJ}}(r_{ij}) =
\varepsilon_{ij} \,
\left [ \left (\frac{r^{{\rm min}}}{r_{ij}} \right )^{12}-2\left (\frac{r^{{\rm min}}}{r_{ij}}\right )^6 \right ] \ ,
\label{Eq. Lennard-Jones}
\end{equation}
where $\varepsilon_{ij}=\sqrt{\varepsilon_i \, \varepsilon_j}$ and $r^{{\rm min}} = (r^{{\rm min}}_i+r^{{\rm min}}_j)/2$.
The corresponding parameters for all atoms in the system have been taken from Refs.~\cite{Filippova2015, Mayo1990} and are summarized in Table~\ref{Table:van_der_Waals}.

\begin{table}[t!]
\caption{Parameters of the Lennard-Jones potential describing the van der Waals interaction between atoms of the deposit and the substrate. $\varepsilon$ is the depth of the potential energy well and $r^{{\rm min}}$ is the interatomic distance corresponding to the potential energy minimum.}
\centering
\begin{tabular}{p{1.2cm}p{3cm}p{2cm}p{1cm}}
\hline
Atom	& $\varepsilon$ (kcal/mol) & $r^{{\rm min}}/2$~(\AA) & Ref. \\
\hline
	Fe    & 16.73   & 1.13   &  \cite{Filippova2015} \\
	C     & 0.095   & 1.95   &  \cite{Mayo1990} \\
	O     & 0.096   & 1.76   &  \cite{Mayo1990}  \\
	Si    & 0.310   & 2.14   &  \cite{Mayo1990} \\
	H     & 0.046   & 0.225  &  \cite{Mayo1990}    \\
		\hline
\end{tabular}
\label{Table:van_der_Waals}
\end{table}

\begin{figure*}[t!]
\centering
	\includegraphics[width=0.9\textwidth]{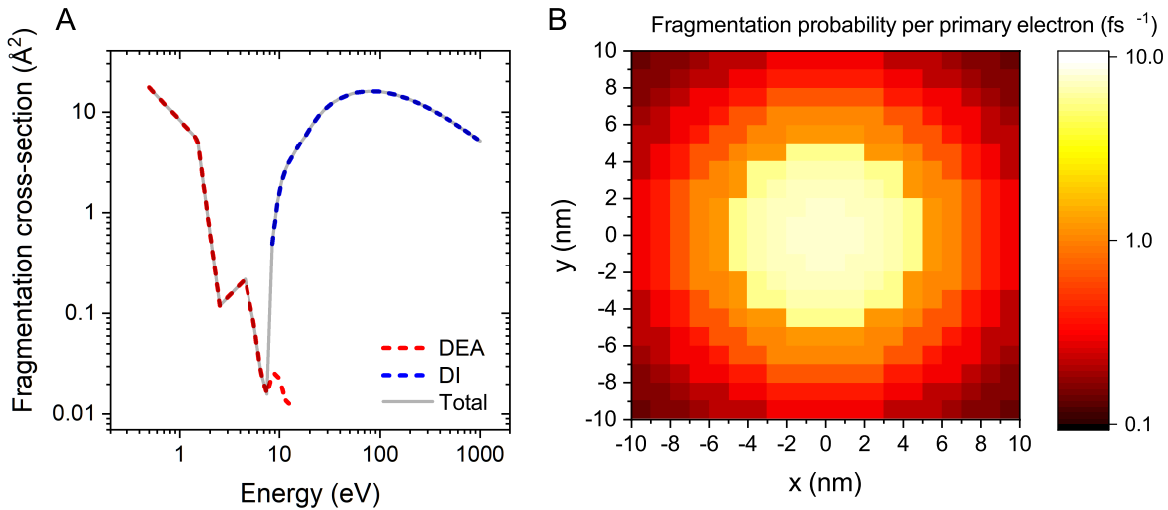}%
	\caption{Panel~A: Total electron impact fragmentation cross section of Fe(CO)$_5$ (solid line), including the dissociative electron attachment (DEA) and dissociative ionization (DI) contributions (dashed lines). Panel~B: Spatial distribution of the fragmentation probability of Fe(CO)$_5$ on SiO$_2$-H, irradiated with a 1~keV electron beam focused within the beam spot of radius $R_{\rm sim} = 5$~nm. The presented probability distribution has been calculated for the unit primary electron flux density of 1~\AA$^{-2}$fs$^{-1}$.}
\label{Fig:Cross-section}
\end{figure*}

The electron-induced fragmentation cross section of Fe(CO)$_5$ as a function of electron energy, $\sigma_{\rm frag}(E)$, is shown in Fig.~\ref{Fig:Cross-section}A. The total fragmentation cross section (solid gray line) is summed up from the cross sections of dissociative electron attachment (DEA) and dissociative ionization (DI). DEA and DI are the main fragmentation channels for precursor molecules at projectile electron energies below and above the ionization threshold of the molecule, respectively \cite{Thorman2015}. The cross section for DEA to Fe(CO)$_5$ has been taken from Ref.~\cite{Allan_PCCP2018}.
Partial cross sections of ionization resulting in the formation of different positively charged fragments were measured experimentally for several precursor molecules in Refs.~\cite{Engmann2013, Thorman2015, Wnorowski2012}. The cited studies showed that almost every ionizing collision of a projectile electron with a molecule leads to fragmentation of the latter. Hence the DI cross section for a Fe(CO)$_5$ molecule has been approximated by its total ionization cross section. Since there are no published data on the total electron-impact ionization cross section of Fe(CO)$_5$, it has been approximated using the additivity rule as the sum of ionization cross sections of a Fe atom \cite{Bartlett2002} and a CO molecule \cite{Hwang1996_BEB}, multiplied by five.

The energy resolved fragmentation cross section $\sigma_{\rm frag}(E)$ has been used to calculate the space resolved fragmentation probability $P(x,y)$ per unit time \cite{DeVera2020}:
\begin{eqnarray}
P(x,y) &=& \sigma_{\rm frag}(E_0) J_{\rm PE}(x,y,E_0) \nonumber \\ 
&+& \sum_i \sigma_{\rm frag}(E_i) J_{\rm SE/BSE}(x,y,E_i)  \ .
\label{Eq. Frag_Probability_total}
\end{eqnarray}
Here $E_i < E_0$ is the electron energy discretized in steps of 1~eV, and $J_{\rm PE/SE/BSE}$ is the flux density of PEs, SEs and BSEs, respectively. Further details on the evaluation of the probability $P(x,y)$ can be found in Ref.~\cite{DeVera2020}.

In the present study we have employed the distribution of electrons calculated previously \cite{DeVera2020} using the track-structure Monte Carlo code SEED and considered a cylindrical PE beam with a radius of 5~nm and energy of 1~keV. Figure~\ref{Fig:Cross-section}B shows the spatial distribution of the fragmentation probability of Fe(CO)$_5$ per one primary electron, calculated according to Eq.~(\ref{Eq. Frag_Probability_total}).
The calculated probability is based on the unit PE flux of 1~\AA$^{-2}$fs$^{-1}$ and is written in the tabulated form for a 20~nm~$\times$~20~nm grid covering the whole simulation box.

The amount of energy deposited into the system upon the interaction with electrons has been determined from the results of DFT calculations carried out in Ref.~\cite{Lacko_EPJD2015}. In the cited paper the energy difference between the ground state of a neutral Fe(CO)$_5$ molecule and a Fe(CO)$_4^+$ fragment, produced due to the emission of a CO ligand from the ionized Fe(CO)$_5^+$, was found equal to 8.7~eV for Fe--C$_{\rm eq}$ bonds and 9.2~eV for Fe--C$_{\rm ax}$ bonds.

In the present study, a layer of Fe(CO)$_5$ with the dimensions of 20~nm~$\times$~20~nm has been created using MBN Studio \cite{Sushko2019}, optimized using the velocity quenching algorithm, softly deposited onto the SiO$_2$-H substrate and thermalized at 300~K for 0.3~ns using the Langevin thermostat with a damping time of 0.2~ps \cite{Prosvetov2021}.
The simulations have been performed using the Verlet integration algorithm with a time step of 1~fs and reflective boundary conditions.
The linked cell algorithm \cite{Solovyov2012} with a cell size of 10~\AA~has been employed for the more efficient evaluation of interatomic interactions between particles of the system.

The FEBID process consists of two iterating phases, namely irradiation with a pulsed electron beam and replenishment of the precursors, which are repeated multiple times \cite{Huth2012}. The irradiation phase lasts for a period called dwell time $\tau_d$, which varies in FEBID experiments from sub-microseconds to sub-milliseconds \cite{Utke2008, Weirich2013}. As the realistic experimental time scale for $\tau_d$ is challenging for all-atom MD, the simulated PE fluxes $J$ (and hence PE beam currents $I$) have been rescaled to match the same PE fluence (i.e. the same number of PE per unit area and per dwell time) as in experiments \cite{Sushko2016}. The correspondence of simulated results to experimental ones is established through the correspondence of the PE fluence per dwell time in simulations and experiments \cite{Sushko2016}. This approach is valid in the case when different fragmentation events occur independently and do not induce a collective effect within the system. In this case the irradiation conditions for the adsorbed precursor molecules are the same in simulations and in experiments. The aforementioned correspondence condition gives
\begin{equation}
I_{{\rm sim}} =
I_{{\rm exp}} \, \lambda \, \frac{S_{{\rm sim}}}{S_{{\rm exp}} } =
I_{{\rm exp}} \, \lambda \, \frac{R_{{\rm sim}}^2}{R_{{\rm exp}}^2 } \ ,
\end{equation}
\begin{equation}
\lambda = \frac{\tau_d^{{\rm exp}}}{\tau_d^{{\rm sim}}} \ ,
\end{equation}
where $S_{{\rm exp}}$ and $S_{{\rm sim}}$ are the electron beam cross sections used in experiments and simulations, respectively; $R_{{\rm exp}}$ and $R_{{\rm sim}}$ are the corresponding beam spot radii.

The following experimental irradiation parameters have been used in the simulations: electron current $I_{{\rm exp}} =$ 1, 2 and 4~nA, and the beam spot radius $R_{{\rm exp}} = 10$~nm. The experimental dwell time of a single irradiation cycle has been set equal to $\tau_d^{{\rm exp}} = 160$~$\mu$s according to Ref.~\cite{Gavagnin2014}. Following the earlier IDMD-based studies of FEBID \cite{Sushko2016, DeVera2020, Prosvetov2021} each irradiation phase has been simulated for the time $\tau_d^{{\rm sim}} = 10$~ns.

The replenishment phase has been simulated to reproduce the system's physical state before the next irradiation cycle.
For specific values of pressure and temperature of a precursor gas, the system's state after the replenishment is characterized by the number of desorbed fragments and the spatial distribution of newly adsorbed precursor molecules; see the detailed description in Refs.~\cite{Sushko2016, Prosvetov2021}.
As a first step of the simulated procedure, weakly bound precursor molecules and fragments have been removed from the system by an external force field. Then, a new layer of precursor molecules has been created at a certain distance above the surface, optimized and deposited upon the substrate. New precursor molecules have been deposited within the circular area with a radius of 8~nm. The selected area covers the PE beam spot and a halo of SEs, while preventing the accumulation of non-fragmented molecules along the simulation box boundaries where the fragmentation probability is very low (see Fig.~\ref{Fig:Cross-section}B).
The average density of newly added molecular layers over the whole simulation box has been set equal to 4 and 6 molecules per nm$^2$. The corresponding surface density within the beam spot area is approximately equal to 6.2 and 8.2 molecules per nm$^2$, respectively.
Finally, the system has been thermalized for 0.5~ns at the end of each replenishment phase, followed by the next cycle of irradiation.

The IDMD simulations of the FEBID process for Fe(CO)$_5$ have been performed for 10 irradiation-replenishment cycles with the total duration of the irradiation phase of 100~ns.

\section{Results and discussions}
\label{Results}

In this study, the operating regime of the FEBID process is defined by the fraction of adsorbed precursor molecules undergoing fragmentation during a specific dwell time.
If at specific irradiation and replenishment conditions, a non-zero fraction of adsorbed precursors in the irradiated area remains intact by the end of a dwell time, FEBID operates in the ELR.
Dissociation of all the adsorbed precursor molecules within a specific dwell time corresponds to the PLR.
As discussed in the previous section, precursor molecules added during each subsequent replenishment phase are deposited primarily in the PE beam spot area to avoid the accumulation of non-fragmented molecules near the simulation box boundaries. Therefore, in what follows, we focus mainly on the analysis of irradiation-driven transformations of the deposit occurring within the PE beam spot area.

\begin{figure}[h!]
\centering
	\includegraphics[width=0.48\textwidth]{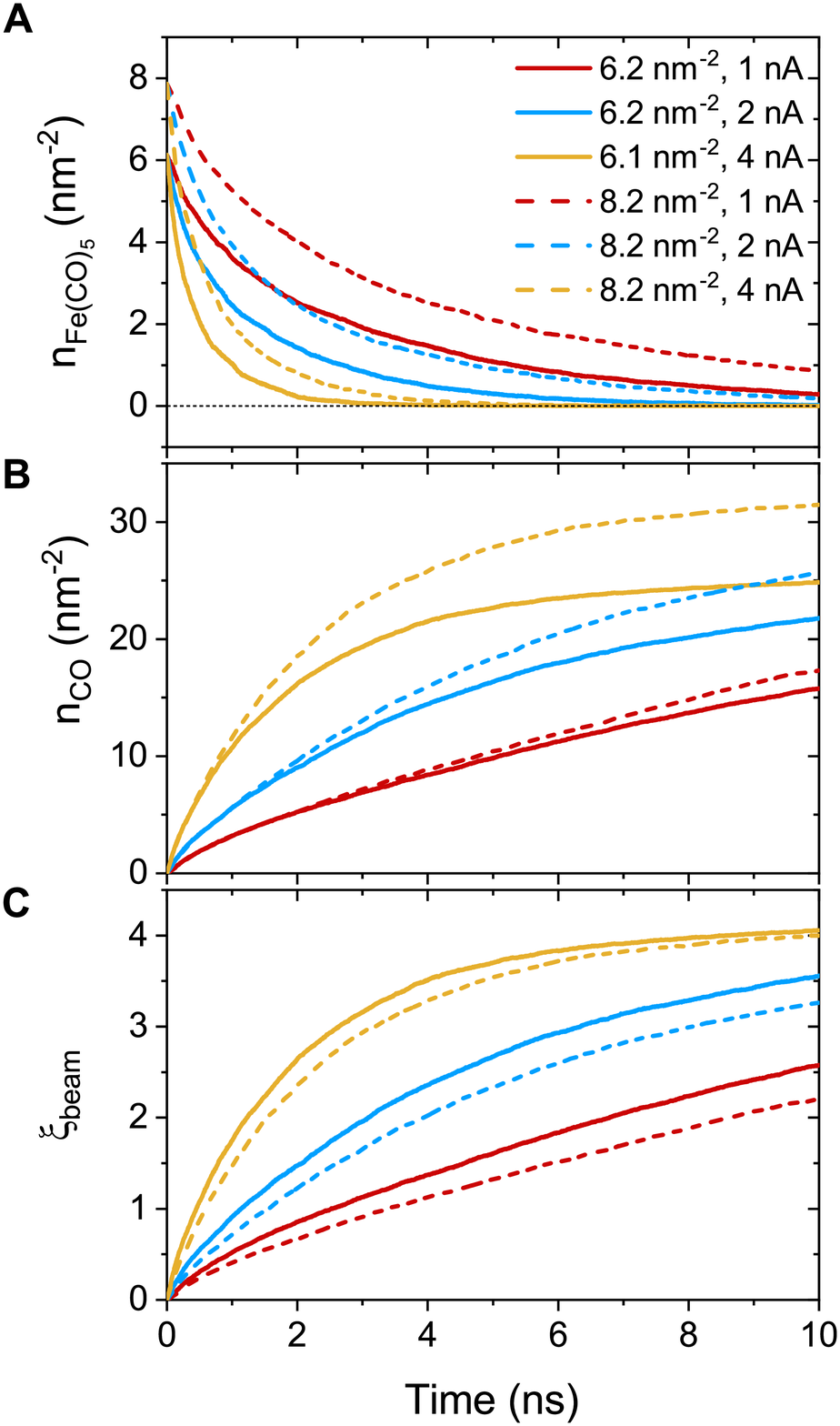}%
	\caption{Variation of the concentration of adsorbed precursor molecules (panel~\textbf{A}) and the concentration of released CO ligands (panel~\textbf{B}) during one irradiation phase with the uniform distribution of adsorbed precursor molecules and the fragmentation probability corresponding to the PE beam spot area. Panel~\textbf{C} shows the variation of an average degree of precursor fragmentation $\xi$, Eq.~(\ref{eq:xi}), in the course of irradiation.}
\label{Fig:PrecursorEvolution}
\end{figure}

Figure~\ref{Fig:PrecursorEvolution}A and Fig.~\ref{Fig:PrecursorEvolution}B show, respectively, the evolution of the concentration of Fe(CO)$_5$ molecules and released CO ligands in the system during one irradiation phase
for the chosen values of electron beam current and precursor surface density in the PE beam spot area. During a 10~ns-long irradiation phase, the concentration of precursors decreases (Fig.~\ref{Fig:PrecursorEvolution}A) while the concentration of released CO ligands increases (Fig.~\ref{Fig:PrecursorEvolution}B) due to the electron-induced fragmentation.
As described in Ref.~\cite{Sushko2016} and also in Section~\ref{sec:Methodology} above, each subsequent irradiation phase is preceded by a replenishment phase, during which (i) new precursor molecules are added to the simulation box and (ii) CO ligands and other volatile molecular fragments are removed from the system.

Figure~\ref{Fig:PrecursorEvolution}A demonstrates that at $I_{{\rm exp}} = 1$~nA (red curves) the concentration of non-fragmented Fe(CO)$_5$ molecules at the end of a 10~ns-long irradiation phase is above zero, indicating the ELR.
Irradiation with the beam current $I_{{\rm exp}} = 4$~nA (orange curves) leads to the rapid fragmentation of all precursor molecules adsorbed on the surface over the time much shorter than the dwell time, indicating the PLR.
Irradiation with the beam current $I_{{\rm exp}} = 2$~nA (blue curves) corresponds to an intermediate regime, in which the FEBID process operates either in the ELR or the PLR depending on the surface density of adsorbed precursors. We note that the concentration of intact precursor molecules at the end of the irradiation phase depends on the chosen dwell time $\tau_d$.
Therefore, by considering shorter or longer dwell times (while keeping all other irradiation parameters unchanged), the  concentration of intact precursors can be higher or lower than the values shown in Fig.~\ref{Fig:PrecursorEvolution}A. A variation of dwell time would affect the range of beam currents corresponding to ELR and PLR. In the present study the fixed value of dwell time, $\tau_d = 10$~ns, has been set for all simulations.

In order to further classify different irradiation and replenishment conditions considered in the present study, let us introduce a dimensionless parameter
\begin{equation}
\xi(t) = \frac{n_f}{n_0}
\label{eq:xi}
\end{equation}
that defines an average degree of precursor fragmentation during irradiation. Here, $n_f$ is the number of precursor fragmentation events during the dwell time and $n_0$ is the number of precursors adsorbed on the surface during the replenishment phase. In the present study we do not consider the dissociation of C--O bonds; hence the number of precursor fragmentation events is approximated by the number of released CO ligands. According to this definition, $\xi$ varies from zero (i.e. no precursor fragmentation) to the total number of ligands in an intact molecule, i.e. $\xi = 5$ for Fe(CO)$_5$. Figure~\ref{Fig:PrecursorEvolution}C shows the evolution of $\xi$ during one irradiation phase for the uniform electron irradiation with the precursor fragmentation probability and precursor surface density corresponding to the beam spot area.

\begin{table*}[t!]
\caption{Characterization of the deposits in different spatial regions of the simulated system. $I_{{\rm exp}}$ is the experimental beam current. $\bar{n}$ stands for the average surface density of Fe(CO)$_5$ precursors added during each replenishment phase. $n_{\rm beam}$, $n_{\rm halo}$ and $n_{\rm per}$ stand for the precursor surface densities in the PE beam spot area, in the 3~nm wide halo surrounding the beam spot, and in the peripheral region at distances above 8 nm from the beam axis, respectively. The degrees of fragmentation $\xi_{\rm beam}$, $\xi_{\rm halo}$ and $\xi_{\rm per}$ have been calculated in additional simulations of the uniform electron irradiation for dwell time $\tau_d = 10$~ns with the precursor fragmentation probability and the precursor surface density corresponding to the respective spatial regions.
 }
\centering
\begin{tabular}{p{1.2cm}|p{1.6cm}|p{1.8cm}p{1.3cm}|p{1.8cm}p{1.3cm}|p{1.8cm}p{1.3cm}}
\hline
$I_{{\rm exp}}$ (nA)& $\bar{n}$ (nm$^{-2}$) & $n_{\rm beam}$ (nm$^{-2}$)	&  $\xi_{\rm beam}$  & $n_{\rm halo}$ (nm$^{-2}$)	&  $\xi_{\rm halo}$   & $n_{\rm per}$ (nm$^{-2}$)	&  $\xi_{\rm per}$   \\
 		\hline
1   & 4.0	&   6.2   &    2.6  &  7.5  &  0.5  &  1.9  &  0.6 \\
2   & 4.0	&	6.2   &    3.5  &  5.5  &  1.0  &  0.7  &  1.2  \\
4   & 4.0	& 	6.2   &    4.1  &  4.6  &  1.8   &  0.4  & 2.1  \\
	   	\hline
1   & 6.0	& 	8.2   &    2.2  &  7.8  &  0.6  &  1.6  & 0.6  \\
2   & 6.0	&	8.2   &    3.3  &  5.9  &  1.0  &  0.5  & 1.3  \\
4   & 6.0	&	8.2   &    4.0  &  5.3  &  1.7  &  0.5  & 2.0  \\
		\hline
\end{tabular}
\label{Table:FEBID_Regimes}
\end{table*}

As it is shown in Fig.~\ref{Fig:Cross-section}B, the probability of precursor fragmentation is nearly constant in the PE beam spot area and decreases rapidly at larger distances from the geometrical center of the beam. The radial distribution of the concentration of Fe(CO)$_5$ molecules added at each replenishment phase has a similar profile. The concentration of newly added precursors is maximal in the beam spot area while it is several times smaller in the region surrounding the beam spot. In order to account for the radial variation of the precursor concentration and fragmentation probability, the degree of fragmentation $\xi$ has been evaluated in three spatial regions: (i) the PE beam spot with a radius of 5~nm; (ii) a diffusive ``halo'' region around the beam spot with a radius from 5 to 8 nm; and (iii) a peripheral region located at distances greater than 8~nm from the beam spot axis. The average precursor fragmentation rate in these regions is equal to 8.5, 1.1 and 0.5 fs$^{-1}$, respectively, for the unit PE flux density $J_0 = 1$~\AA$^{-2}$fs$^{-1}$ (see Fig.~\ref{Fig:Cross-section}B).
The average precursor concentration $n$ in the aforementioned regions, evaluated at the beginning of each irradiation phase, has been used as input for complementary simulations carried out to estimate the values of $\xi$. These simulations describe the uniform electron irradiation with the precursor fragmentation probability and the precursor surface density corresponding to the respective spatial regions. The obtained values of $n$ and $\xi$ in the beam spot, halo and peripheral regions are listed in Table~\ref{Table:FEBID_Regimes} with the subscripts ``beam'', ``halo'' and ``per'', respectively.

\begin{figure*}[t!]
\centering
	\includegraphics[width=0.85\textwidth]{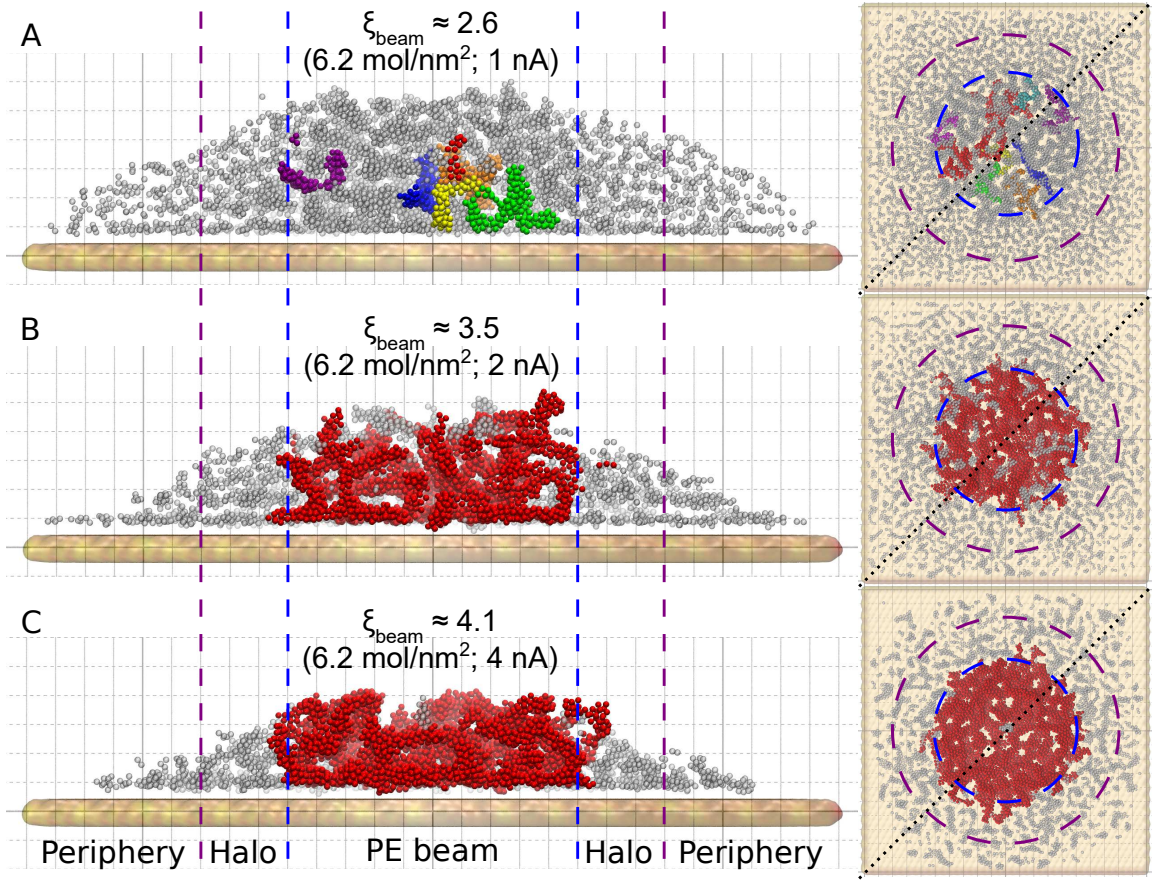}%
	\caption{Snapshots of the IDMD simulations of the FEBID process for Fe(CO)$_5$ after ten irradiation-replenishment cycles (left column -- side view on diagonal cross sections indicated by dotted lines, right column -- top view). The snapshots correspond to the maximal adsorbed precursor surface density in the beam spot area, $n_{\rm beam} = 6.2$~mol/nm$^2$. Panels A, B and C correspond to the electron current $I_{\rm exp}$= 1, 2 and 4~nA, respectively. Only iron atoms are shown for clarity. Topologically disconnected metal clusters consisting of more than 100 iron atoms are shown in different colors. Smaller clusters containing less than 100 iron atoms are shown in grey. Boundaries of the PE beam, halo and peripheral regions are indicated by dashed lines in the left column and by circles in the right column. Grid line spacing equals to 1~nm in all dimensions.}
\label{Fig:Snapshot-4mol}
\end{figure*}

Figure~\ref{Fig:Snapshot-4mol} shows the atomistic structure of the deposits formed after ten irradiation-replenishment cycles at the indicated conditions. Only iron atoms are shown for clarity. Different colors indicate topologically disconnected metal structures, as follows from the analysis of the system's molecular topology.
The deposit grown at electron current $I_{{\rm exp}} = 1$~nA (Fig.~\ref{Fig:Snapshot-4mol}A) is characterized by disconnected metal clusters of a small size. Due to the low precursor dissociation rate (and hence, a relatively small degree of fragmentation, $\xi_{\rm beam} \approx 2.6$) many CO ligands remain attached to the iron clusters, thus preventing their agglomeration and interconnection into larger islands. At higher values of electron current the clusters start to merge into larger dendrite-like structures. In the case of $I_{{\rm exp}} = 4$~nA ($\xi_{\rm beam} \approx 4.1$; see Fig.~\ref{Fig:Snapshot-4mol}C), iron clusters have coalesced into one bigger structure within the first few irradiation cycles. The lateral size of the resulting structure (shown by the red color) corresponds to the PE beam spot area. In the case of $I_{{\rm exp}} = 4$~nA small iron clusters in the halo and peripheral regions are formed at the conditions corresponding to a low degree of fragmentation.
The size and morphology of clusters formed outside the beam spot area at $I_{{\rm exp}} = 4$~nA are similar to those formed in the PE beam area in ELR at $I_{{\rm exp}} = 1$~nA. However, the deposit's height in the peripheral region is much smaller than in the PE beam spot region because a much smaller number of precursors have been adsorbed into that region.

\begin{figure*}[bt!]
\centering
	\includegraphics[width=0.9\textwidth]{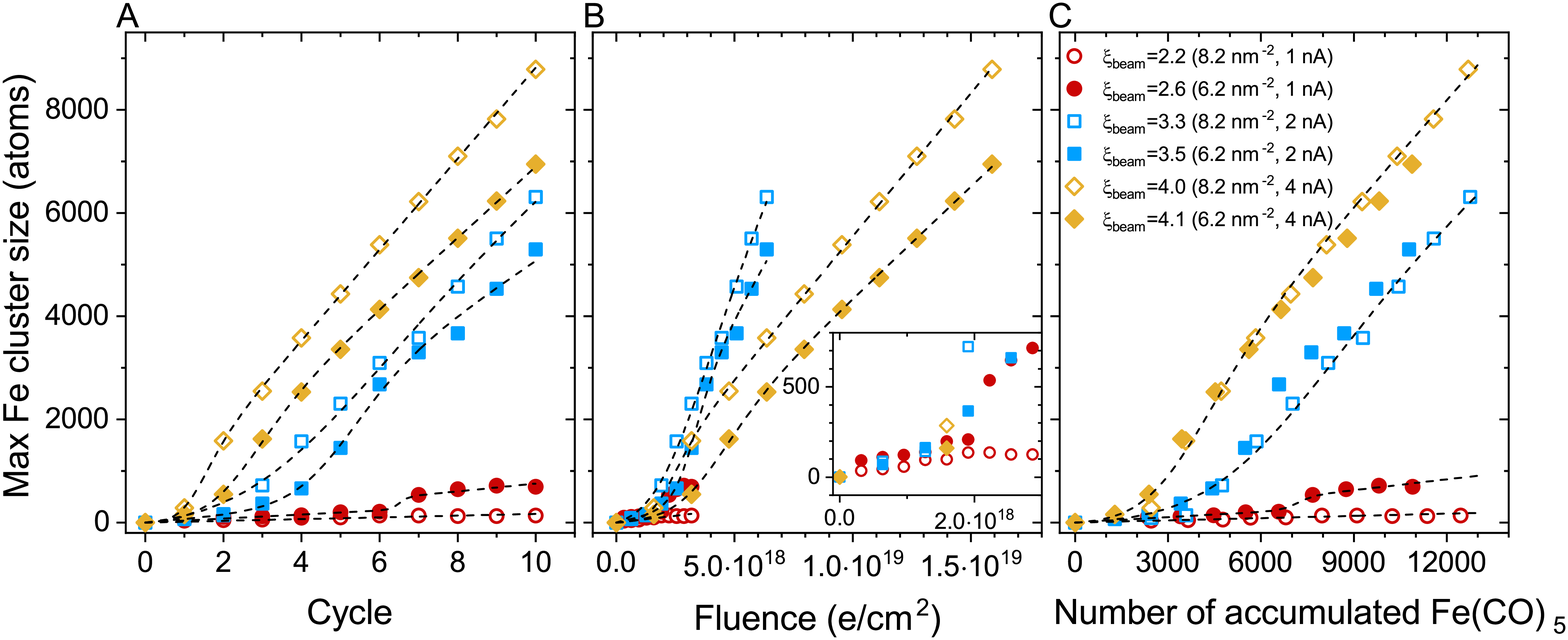}%
	\caption{The maximal number of iron atoms in the largest cluster, $N_{\rm max}$, as a function of FEBID cycle (panel~A), electron fluence (panel~B) and the number of adsorbed Fe(CO)$_5$ precursors (panel~C) for different irradiation-replenishment conditions considered in this study. Each $N_{\rm max}$ value has been evaluated at the end of each consecutive FEBID cycle. The dashed lines are plotted to guide the eye. The inset in panel~B shows a zoomed-in region of low electron fluence.}
\label{Fig:Precursors_FeClusterSize}
\end{figure*}

The evolution of the maximal iron cluster size, $N_{\rm max}$, for consecutive FEBID cycles is shown in Fig.~\ref{Fig:Precursors_FeClusterSize}A.
The cluster growth follows different trends depending on the FEBID regime, i.e. irradiation and replenishment conditions. In general, the deposit's growth can be characterized by three stages: (1) nucleation of Fe(CO)$_5$ molecular fragments and the formation of small iron-containing clusters, (2) coalescence of isolated clusters in a bigger structure, which corresponds to a fast increase of the number of iron atoms in such a structure, and (3) a steady-state linear growth of the resulting structure. All these stages can be seen for the case of $I_{{\rm exp}} = 2$~nA (blue symbols). The evolution of the iron-containing deposit at this beam current is similar to our earlier results on the FEBID of Pt(PF$_3$)$_4$ precursors \cite{Prosvetov2021}. At lower electron current ($I_{{\rm exp}} = 1$~nA, red symbols), the coalescence is significantly suppressed resulting in a small increase of the cluster size up to several hundreds of iron atoms. In the case of $I_{{\rm exp}} = 4$~nA (orange symbols), the nucleation of isolated clusters and their coalescence occur within the first 2-3 FEBID cycles.

The cluster aggregation process is governed by an interplay of two parameters. The surface density of precursors in the irradiated area determines the number of metal atoms in the system and the corresponding interatomic distances. The other parameter is the rate of precursor fragmentation which is proportional to electron flux density and the fragmentation cross section, see Eq.~(\ref{Eq. Frag_Probability_total}). The cumulative effect of these parameters on the deposit's growth can be correlated to the degree of fragmentation $\xi$. In the case of a low degree of precursor fragmentation ($\xi_{\rm beam} \approx 2.2 - 2.6$; red symbols), the maximal iron cluster size saturates, and the metal grains do not grow further as the FEBID process continues. At a higher degree of precursor fragmentation ($\xi_{\rm beam} \approx 3.3 - 4.1$), a morphological transition from isolated metal grains to a bigger metal structure occurs. Metal clusters merge into a bigger structure, which grows linearly during subsequent FEBID cycles. Comparing the evolution of $N_{\rm max}$ for the considered values of $\xi$ (i.e. for the irradiation and replenishment parameters considered in the present study) permits to estimate a threshold value of $\xi_{\rm beam}$ for the coalescence of isolated iron clusters. According to the results shown in Fig.~\ref{Fig:Precursors_FeClusterSize}A the threshold value lies within the range $\xi_{\rm beam} \sim 2.6 - 3.3$.

To analyze the impact of precursor surface density and the fragmentation rate on the cluster growth, the maximal iron cluster size $N_{\rm max}$ is also plotted as a function of accumulated electron fluence $F$ (Fig.~\ref{Fig:Precursors_FeClusterSize}B)  and the accumulated number of adsorbed precursor molecules $N_{\rm p}$ (Fig.~\ref{Fig:Precursors_FeClusterSize}C).

In our earlier study on the FEBID of Pt(PF$_3$)$_4$ \cite{Prosvetov2021} the merging of small clusters into a bigger structure was simulated, and the dependence of $N_{\rm max}$ on $N_{\rm p}$ was fitted by a function containing a linear and sigmoid terms. In the present study, the deposit's growth has been simulated up to the next stage corresponding to the linear growth of the nanostructure after the coalescence of separated clusters. The three stages of the deposit's growth can be taken into account by fitting the $N_{\rm max}(N_{\rm p})$ dependence obtained in the simulations by the following function:
\begin{equation}
N_{\rm max} = k_1 \, N_{\rm p} + \frac{k_2 \, N_{\rm p} + C }{1+\exp{\left(- \frac{N_{\rm p} - N_{\rm tr}}{\Delta N} \right)}} \ .
\label{Eq. MaxClusterSizeSigmoid}
\end{equation}
Here $k_1$ is a linear coefficient of the initial growth of separate clusters.
$\Delta N$ is an interval of $N_{\rm p}$ values within which the morphological transition from isolated metal islands to a bigger nanostructure takes place. $N_{\rm tr}$ is the sigmoid's midpoint corresponding to the fastest growth rate during the morphological transition. The coefficient $k_2$ defines the linear growth of the metal nanostructure after the coalescence of isolated clusters into a bigger structure. When the morphological transition has not been observed, the parameter $C$ defines the maximal cluster size for the saturated growth of separate metal grains. When the morphological transition takes place, $C$ stands for an offset of the asymptotic linear growth after merging of the clusters.

The fitting parameters entering Eq.~(\ref{Eq. MaxClusterSizeSigmoid}) are summarized in Table~\ref{Table:ClusterSizeFit}. In the case of $\xi_{\rm beam}=2.2$ only the first (linear) stage of the deposit's growth takes place at the considered irradiation and replenishment conditions. At $\xi_{\rm beam} = 2.6$ the coalescence process occurs, but only for several iron clusters. In the case of $\xi_{\rm beam} = 3.3$ and 3.5 ($I_{\rm exp} = 2$~nA), all three stages of the deposit's growth are present. At a higher fragmentation rate ($\xi_{\rm beam} = 4.0$ and 4.1) the merging process happens much faster, and the initial nucleation and linear growth are completed within the first 1-2 FEBID cycles.

\begin{table}[h!]
\caption{Degree of precursor fragmentation $\xi_{\rm beam}$ and the parameters of Eq.~(\ref{Eq. MaxClusterSizeSigmoid}) providing a fit to the data obtained in the simulations.
	$k_1$ is the coefficient of the initial linear growth of isolated metal clusters. $\Delta N$ is an interval of $N_{\rm p}$ values within which the morphological transition from isolated metal islands to a bigger nanostructure takes place. $N_{\rm tr}$ is the number of adsorbed precursors at the point of a morphological transition from isolated metal islands to a single metal nanostructure. $k_2$ is the coefficient of the linear growth of the metal nanostructure formed after the coalescence of isolated clusters. The positive value of $C$ indicates the maximal cluster size after reaching saturation. The negative values of $C$ define an offset of the asymptotic linear growth after coalescence of the clusters. Missing values corresponds to the irradiation regimes and intervals of parameters which have not been covered by the performed simulations.}
\centering
\begin{tabular}{p{1.2cm}p{1.2cm}p{1.2cm}p{1.2cm}p{1.2cm}p{1.2cm}}
\hline
 $\xi_{\rm beam}$  & $k_1$  &  $N_{\rm tr}$  & $\Delta N$   & $k_2$  & $C$   \\
 		\hline
 2.2  &  0.01  &  --     &  --   &  --    &  --       \\
 2.6  &  0.03  &  7500   & 132    & 0     & 350   \\
 3.3  &  0.04  &  4400   & 683   &  0.6  & -2070 \\
 3.5  &  0.04  &  4759   & 724    & 0.6  &  -1209     \\
 4.0  &      &  2808     & 459     & 0.8  & -883  \\
 4.1  &      &  2453     & 677     & 0.7   & -359 \\
	   	\hline
\end{tabular}
\label{Table:ClusterSizeFit}
\end{table}

Figures~\ref{Fig:Precursors_FeClusterSize}A and \ref{Fig:Precursors_FeClusterSize}C demonstrate also that a higher degree of fragmentation $\xi$ corresponds to the faster process of cluster nucleation, coalescence and growth. The main requirement for coalescence of clusters during the FEBID process is the contact of separate clusters with dangling bonds created after the dissociation of Fe(CO)$_5$ molecules. In the present case study of Fe(CO)$_5$, metal clusters grow  via the formation of bonds between iron atoms. When the degree of fragmentation is low, remaining ligands covering the metal core prevent the approaching of separate metal clusters to a distance of the Fe--Fe bond length. High values of $\xi$ correspond to a smaller number of ligands attached to metal clusters. This leads to a higher probability that separate clusters approach each other at short distances at which the metal bond between Fe atoms is formed.

The coalescence of metal clusters in the grown deposit may govern the deposit's properties, such as thermal and electrical conductivity. The interconnection of separate clusters can be correlated with a jump in electrical current through the deposit observed in time-dependent electrical conductivity measurements during FEBID \cite{Porrati2009, Hochleitner2008}.

\begin{figure*}[t!]
\centering
	\includegraphics[width=0.85\textwidth]{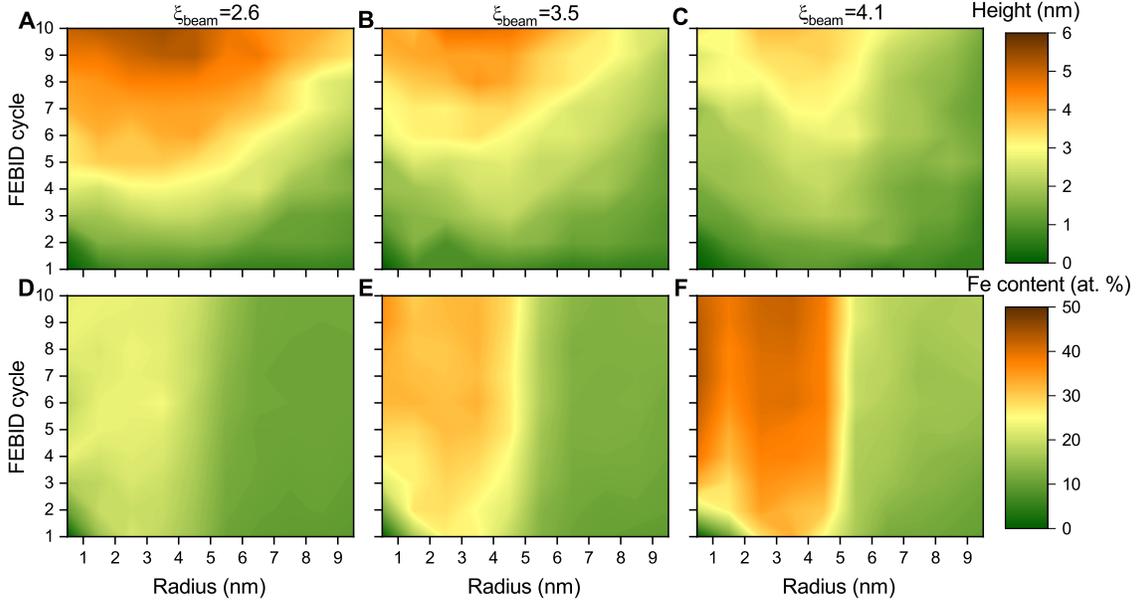}%
	\caption{Radial distribution of the height (panels~A, B, C) and the atomic Fe content (panels~D, E, F) of the grown iron-containing structures as functions of the number of simulated FEBID cycles for electron currents of 1~nA ($\xi_{\rm beam} \approx 2.6$), 2~nA ($\xi_{\rm beam} \approx 3.5$) and 4~nA ($\xi_{\rm beam} \approx 4.1$). }
\label{Fig:FeHeight_Content_Radial}
\end{figure*}

The height and metal content of the deposits grown by FEBID are the key characteristics that can be measured experimentally. These quantities strongly depend on the FEBID regime at which the process operates \cite{Lavrijsen2011, Wachter2014}.
In the present study, the height and relative iron content of the deposited metal structures have been evaluated in concentric bins with a width of 1~nm around the beam axis due to the cylindrical symmetry of the beam and the distribution of adsorbed precursors.
Figure~\ref{Fig:FeHeight_Content_Radial} shows the radial distribution of the maximal height of iron-containing deposits (upper row -- panels~A, B and C) and metal content (lower row -- panels~D, E and F) at consecutive FEBID cycles for $I_{{\rm exp}} = 1$, 2 and 4~nA and the average surface density of adsorbed precursors $\bar{n} = 4$~mol/nm$^2$. The deposit exhibits different growth patterns depending on the degree of fragmentation $\xi$. For $\xi_{\rm beam} \approx 2.6$ (panels~A and D), the deposit starts to grow covering the entire beam spot area (radius $R_{\rm sim} = 5$~nm) with low concentration of iron.
Although the deposit is accumulated in a broad region surrounding the beam spot, the radial distribution of Fe content indicates that the deposit primarily consists of low-fragmented or non-fragmented Fe(CO)$_5$ precursors. For $\xi_{\rm beam} \approx 4.1$ (panels~C and F), the deposit starts to grow near the boundary of the beam spot area and covers the entire beam spot region within a few FEBID cycles. This effect can be related to the diffusion of irradiated molecules in the beam spot area confined by the surrounding precursors. In this case the deposit is characterized by the highest metal content. For $\xi_{\rm beam} \approx 3.5$ (panels~B and E), the deposit's height and metal concentration are between the two limiting cases.

\begin{figure}[tb!]
\centering
	\includegraphics[width=0.45\textwidth]{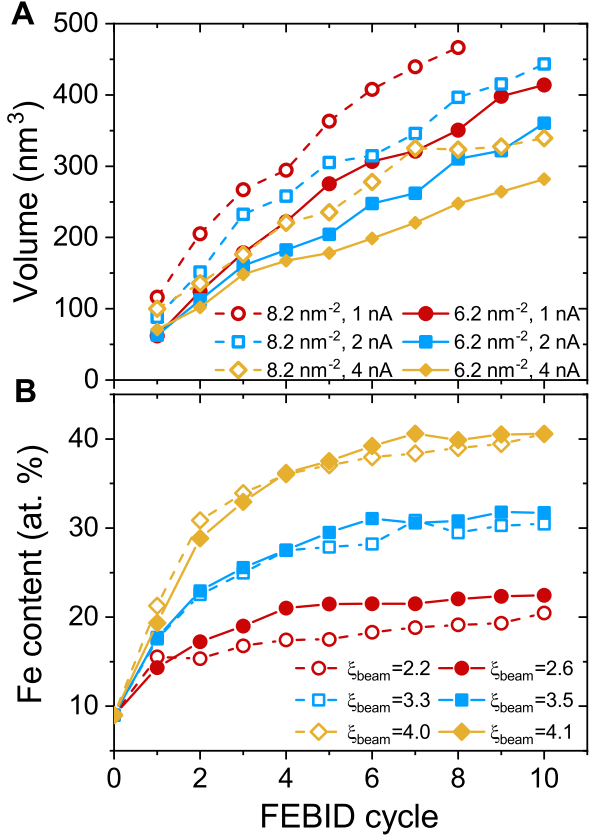}%
	\caption{Evolution of the deposit's volume (panel~\textbf{A}) and the average Fe content in the beam spot area (panel~\textbf{B})
	during ten consecutive FEBID cycles
	for different irradiation-replenishment conditions considered in the present study.} 
\label{Fig:Volume_FeContent_vs_Time}
\end{figure}

Figure~\ref{Fig:Volume_FeContent_vs_Time} shows the evolution of the deposit volume and the average Fe content in the beam spot area for consecutive FEBID cycles. The deposit volume (Fig.~\ref{Fig:Volume_FeContent_vs_Time}A) has been calculated as a sum of radial bin areas multiplied by the height of the grown structure in each bin.
The figure indicates a linear increase in the volume of the deposit from one cycle to another and hence as a function of accumulated electron fluence and the number of precursors.
Increasing the surface density of adsorbed precursors in the beam spot area from 6.2 to 8.2 mol/nm$^2$ (see the solid and open symbols of the same color in Fig.~\ref{Fig:Volume_FeContent_vs_Time}A) for the same electron current results in the formation of taller structures that occupy a larger volume. If the concentration of newly added precursor molecules remains constant, higher electron current leads to the formation of a deposit with a smaller volume.

The characteristics of the deposit correlate with the degree of precursor fragmentation $\xi_{\rm beam}$. The deposit's growth rate, defined by the slope of the dependencies shown in Fig.~\ref{Fig:Volume_FeContent_vs_Time}A, is lower for higher values of $\xi_{\rm beam}$. This indicates that a higher degree of precursor fragmentation leads to the formation of a denser metal deposit. The dependence of the metal concentration on electron current (see Fig.~\ref{Fig:Volume_FeContent_vs_Time}B) indicates that a larger number of ligands detached from precursor molecules during the FEBID process result in higher metal content of the deposits. The maximum values of the Fe content for different values of $\xi_{\rm beam}$ correlate with the evolution of the maximal size of Fe clusters with respect to the number of accumulated precursors (see Fig.~\ref{Fig:Precursors_FeClusterSize}C). In both cases there is a monotonous dependence on $\xi_{\rm beam}$ and there is no significant dependence on the concentration of added precursors. This indicates that the process of initial nucleation and coalescence of metal clusters influences the final metal content of the deposit.
The atomic concentration of Fe saturates at $\sim$20\% for $I_{\rm exp} = 1$~nA and $\xi_{\rm beam} \approx 2.6$. In the case of $I_{{\rm exp}} = 2$~nA and $\xi_{\rm beam} \approx 3.5$, the concentration of iron determined after 10 FEBID cycles saturates at around 30\%. The Fe content of the deposit grown in the regime with the higher degree of fragmentation, $\xi_{\rm beam} \approx 4.1$, increases up to 40\% after 10 simulated FEBID cycles.

A more elaborated analysis of the nanostructure growth at specific experimental conditions for precursor replenishment (i.e. at specific precursor gas pressure and temperature) is a complex task that goes beyond the scope of the present study and can be addressed in follow-up studies.

\begin{figure}[bt!]
\centering
	\includegraphics[width=0.45\textwidth]{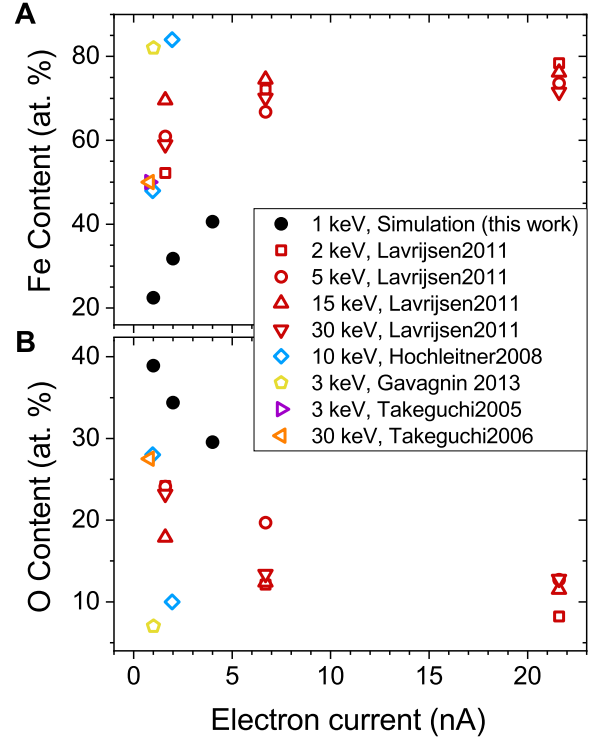}%
	\caption{Fe (panel \textbf{A}) and O (panel \textbf{B}) content in the deposited structures as a function of electron current. Solid black dots correspond to the values obtained in the present study. Open symbols show experimental data from Refs.~\cite{Lavrijsen2011, Hochleitner2008, Gavagnin2013, Takeguchi2005, Takeguchi2008}. Note that the FEBID experiments by Lavrijsen \textit{et al.}~\cite{Lavrijsen2011} were carried out for Fe$_2$(CO)$_9$ precursors.}
\label{Fig:AtContent_vs_Current}
\end{figure}

The experimental values of the Fe content in the FEBID grown deposits vary in a broad range from 40 to 90 at.~\% \cite{Lukasczyk2008, Gavagnin2013, Lavrijsen2011, deTeresa2016}. The highest metal content has been achieved in experiments carried out at ultra-high vacuum conditions, which implies that all volatile molecular fragments are removed from the surface, leaving only the metal-enriched deposit. Figure~\ref{Fig:AtContent_vs_Current} compares atomic content in the deposits obtained in the present study (solid symbols) with experimental data (open symbols) as a function of electron current.

The calculated dependence of the iron content on electron current (Fig.~\ref{Fig:AtContent_vs_Current}A) follows the trend observed in experiments, but the absolute values of the iron content are about 30 at.~\% lower than in experiments at a given electron current. There might be several possible reasons for this discrepancy. First, the energy of PE beam is different in the simulations and experiments. The simulations have been performed for the PE energy of 1~keV, while the experiments \cite{Lavrijsen2011, Hochleitner2008, Gavagnin2013, Takeguchi2005, Takeguchi2008} were carried out at higher PE energies in a broad range from 2 to 30~keV.  As it was shown in Ref.~\cite{DeVera2020}, the yield, energy and spatial distributions of SEs, which play a significant role in the precursor dissociation, vary significantly for different energies of PEs. Another reason may be related to the contribution of neutral dissociation (ND) \cite{Thorman2015}, which is not taken into account in the present study. Experimental measurements of ND cross sections are challenging, leading to the lack of available data that can be used as input for IDMD simulations of the FEBID process.
Finally, the difference between the simulation outcomes and experimental data on the deposit's metal content can also be attributed to a simplified irradiation-chemistry model utilized in the present simulations, which accounts only for the fragmentation of metal-ligand bonds and neglects the irradiation-induced dissociation of CO ligands.
The rigidity of CO groups prevents the release of oxygen atoms from the deposits (see Fig.~\ref{Fig:AtContent_vs_Current}B), which would lead to an increase of Fe atomic content in the remaining deposit.
A more elaborated irradiation chemistry model should be utilized to improve the quantitative agreement with experiments, which accounts for the dissociation of ligands and chemical reactions involving the reactive fragments.
The development of such a model is a task for the future advancement of the computational protocol for IDMD simulations of FEBID.

It should be noted that the observed difference between the Fe content determined in the present simulations and experiments  does not influence the analysis of the nanostructure's formation and growth in the ELR and PLR of the FEBID process.

\section{Conclusions}
\label{Conclusions}

The Irradiation Driven Molecular Dynamics method has been used to study the growth of metal nanostructures at different irradiation and replenishment conditions corresponding to the electron-limited and precursor-limited regimes (ELR, PLR) of the FEBID process. Fe(CO)$_5$, one of the most common FEBID precursors, has been chosen as an illustrative case study. The dependence of the deposit's morphology and metal content on the number of adsorbed molecules and electron current has been investigated on the atomistic level.

It has been shown that the morphology and purity of iron-containing nanostructures grown at different irradiation conditions vary significantly. The average degree of precursor fragmentation $\xi$ ranging from 0 to 5 for Fe(CO)$_5$ has been introduced to classify irradiation and replenishment conditions. The studied combinations of the precursors surface density ($n = 4$ and 6~mol/nm$^2$) and electron current ($I_{\rm exp} = 1, 2$ and 4~nA) correspond to a discrete set of $\xi$ values, which vary in the range $\xi_{\rm beam} \approx 2.2 - 4.1$ in the beam spot area. When at the end of dwell time, electron irradiation causes dissociation of some fraction of precursors adsorbed in the beam spot area (this corresponds to the ELR), the average degree of precursor fragmentation in the beam spot area is low, $\xi_{\rm beam} < 2.6$. The remaining ligands prevent the approaching of metal clusters to a distance required for their coalescence into a bigger structure. In this case, the nanogranular deposit consists of small topologically separated metal grains containing up to several hundreds of iron atoms covered with organic ligands. The grown deposit is characterized by high volume and low metal content (ca. 20 at.$\%$).
In PLR all precursor molecules in the beam spot area undergo fragmentation by the end of dwell time. The corresponding higher degree of fragmentation ($\xi_{\rm beam} > 3.3$) enables the coalescence of separate clusters into a large dendrite-like structure with the size corresponding to the primary electron beam. A smaller number of attached ligands, in this case, leads to a smaller volume and higher Fe content (up to 40 at.$\%$) of the grown deposit.

The calculated dependence of the iron content on electron current follows the experimental trend, but the absolute values of the iron content are about 30 at.\% lower than in experiments at a given electron current. The reason for this discrepancy is likely related to the underestimation of the ligand release from the deposits. This can be caused by neglecting the neutral dissociation channel of precursor fragmentation and using a simplified model of irradiation-induced chemistry with rigid C--O bonds.

It should be stressed that the methodology used for the atomistic simulation of the FEBID process is under continuous development \cite{Sushko2016, DeVera2020, Prosvetov2021}. Most of the physical processes occurring during FEBID can be considered within the atomistic approach. The electron-induced dissociation of ligands and the follow-up chemical reactions will be addressed in a separate study to further advance the computational methodology for atomistic IDMD simulations of the FEBID process.

\section*{Acknowledgments}

The authors acknowledge financial support from the Deutsche Forschungsgemeinschaft (Project no. 415716638) and the European Union's Horizon 2020 research and innovation programme -- the RADON project (GA 872494) within the H2020-MSCA-RISE-2019 call.
Part of this work was supported by COST (European Cooperation in Science and Technology) through the COST Actions TUMIEE (CA17126) and MultIChem (CA20129).
The possibility to perform computer simulations at Goethe-HLR cluster of the Frankfurt Center for Scientific Computing and the Supercomputing Center of Peter the Great Saint Petersburg Polytechnic
University is gratefully acknowledged.


\bibliography{MBN-RC.bib}

\end{document}